\documentclass[useAMS,usenatbib]{mn2e}
\usepackage{graphicx}
\title[GRB090111]{GRB090111: extra soft steep decay emission and peculiar re-brightening}
\author[Margutti et al.]{R. Margutti$^{1,2}$\thanks{E-mail:
raffaella.margutti@brera.inaf.it (RM)}, T. Sakamoto$^{3}$,
G. Chincarini$^{1,2}$, C. Guidorzi$^{4,2}$, J. Mao$^{2,6}$, F. Pasotti$^{2}$,\and D. Burrows$^{5}$,
P. D'Avanzo$^{1}$, S. Campana$^{2}$, S.D. Barthelmy$^{3}$, N. Gehrels$^{3}$ \\
$^{1}$Universita' degli studi Milano Bicocca, P.za della Scienza 3, Milano 20126, Italy\\
$^{2}$INAF Osservatorio Astronomico di Brera, via Bianchi 46, Merate 23807, Italy \\
$^{3}$NASA/Goddard Space Flight Center, Greenbelt, MD 20771, USA \\
$^{4}$Dipartimento di Fisica, Universit\`a di Ferrara, via Saragat 1, I-44100, Ferrara, Italy\\
$^{5}$Department of Astronomy and Astrophysics, Pennsylvania State University, 525 Davey Lab, University Park, PA 16802, USA\\
$^{6}$Yunnan Observatory, Chinese Academy of Sciences, Kunming, Yunnan Province, China \\}
\begin{document}

\date{Accepted 200? Month Day  Received 2009  Month Day; in original form 2009 March 19}

\pagerange{\pageref{firstpage}--\pageref{lastpage}} \pubyear{2009}

\maketitle

\label{firstpage}

\begin{abstract}
We present a detailed study of GRB\,090111, focusing on its extra soft power-law
photon index $\Gamma>5$ at the very steep decay phase emission (power-law index $\alpha=5.1$, 
steeper than 96\% of GRBs detected by \emph{Swift}) and the following peculiar X-ray 
re-brightening. Our spectral analysis supports the hypothesis of a comoving Band spectrum with 
the the peak of the $\nu F_{\nu}$ spectrum evolving with time to lower values: a
period of higher temporal variability in the 1-2 keV light-curve ends when the $E_{\rm{peak}}$
evolves outside the energy band.
The X-ray re-brightening shows extreme temporal properties when compared to a homogeneous sample of 
82 early flares detected by Swift. While an internal origin cannot be excluded, we show these properties 
to be consistent with the energy injection in refreshed shocks produced by slow shells colliding 
with the fastest ones from behind, well after the internal shocks that are believed to give rise to 
the prompt emission have ceased.
\end{abstract}


\section{Introduction}
The unprecedented fast re-pointing capability of  Swift \citep{Gehrels04}
has ushered in a new era in the study of Gamma Ray Bursts (GRB) sources.
A canonical picture of the X-ray afterglow light-curve emerged (see e.g. 
\citealt{Nousek06}), 
with five different components describing the overall structure observed 
in the majority of events: an initial steep decay, a shallow-decay plateau 
phase, a normal decay, a jet-like decay component as well as
randomly occuring flares.

The steep decay phase smoothly connects to the prompt emission (e.g. 
\citealt{Tagliaferri05}), with a typical temporal power law 
decay index between 2 and 4 \citep{Evans09a}: this strongly suggests a
common physical origin. The observed spectral softening  with time 
challenges the simplest version of the most 
popular theoretical model for this phase, the High Latitude 
Emission (HLE) model (\citealt{Fenimore96}; \citealt{Kumar00}): 
according to this scenario, steep decay photons originate from the 
delayed prompt emission from different viewing latitudes of the emitting area 
(\citealt{Zhang07}
) and are expected to lie on a simple power law (SPL) spectral model. 
The $0.3-10\,\rm{keV}$ spectrum of the steep decay phase 
is generally consistent  with the expected SPL behaviour with a typical 
photon index  $\Gamma\sim 2$ (see \citealt{Evans09a}); however,
a careful inspection of the GRBs with the 
best statistics reveals that alternative explanations are required (see e.g. 
\citealt{Zhang09}; \citealt{Qin09}). Deviations from the SPL spectral model are 
therefore of particular interest.

Flares have been found to be a common feature of early X-ray afterglows:
with a typical duration over occurrence time $\Delta t/t \sim0.1$ 
\citep{Chincarini07} and a Band spectrum 
\citep{Band93} reminiscent of the typical spectral shape of the prompt 
emission \citep{Falcone07}, they are currently believed to be related to
the late time activity by the central engine. In spite of 
the growing statistics their origin is still an open issue.

In this paper we analyse and discuss how and if 
the extra soft ($\Gamma>5$) steep decay emission of GRB\,090111 
fits into different theoretical models; particular attention will be 
devoted to the possible link with the detected soft prompt $15-150\,\rm{keV}$
emission. After the steep decay, the GRB\,090111 $0.3-10\,\rm{keV}$ 
light-curve shows a peculiar re-brightening, with extreme properties 
when compared to typical X-ray flares: alternative explanations 
are discussed.
The paper is organised as follows: observations are
described in Sect. \ref{sec:observations}; the details of the 
data analysis are reported in Sect. \ref{sec:analysis}.
Our results are discussed in Sect. \ref{sec:discussion}.
Conclusions are drawn in Sect. \ref{sec:conclusion}.
Uncertainties and upper limits are quoted at the 90\% confidence 
level (c.l.) unless otherwise stated.
\section{Swift observations}
\label{sec:observations}

\begin{figure}
  \centering
  \includegraphics[scale=0.48]{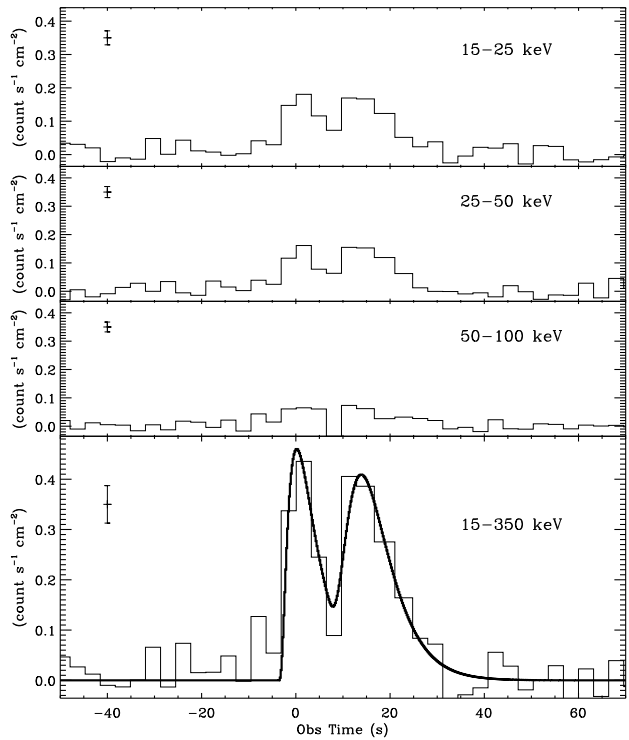}
  \caption{BAT mask weighted light-curve in different energy bands (binning 
    time of 3.2 s). No signal is detected above 100 keV. 
    Bottom panel, solid black line: $15-350\,\rm{keV}$ light-curve best fit. 
    The typical $1\sigma$ error size is also shown in each panel.}
  \label{Fig:BAT}
\end{figure}

\begin{figure}
\centering
    \includegraphics[scale=0.48]{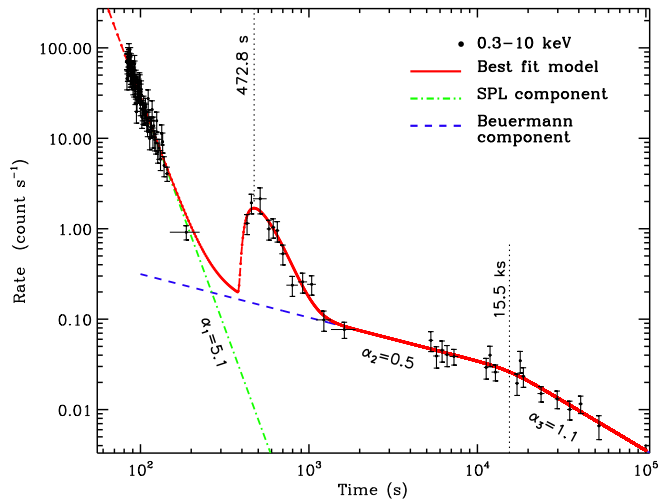}
      \caption{XRT $0.3-10\,\rm{keV}$ light-curve with best fit  superimposed.}
\label{Fig:XRT}
\end{figure}

The Swift Burst Alert Telescope (BAT; \citealt{Barthelmy05}) 
triggered and located GRB\,090111 at 23:58:21 UT on 2009-01-11.
The spacecraft immediately 
slewed to the burst allowing the X-ray Telescope (XRT; \citealt{Burrows05}) 
and the UV/Optical Telescope (UVOT, \citealt{Roming05}) to collect 
data starting $76.6\,\rm{s}$ and $86\,\rm{s}$ after the trigger,
respectively. A refined position was quickly available: 
R.A.(J2000)=$16^{\rm{h}}46^{\rm{m}}42.14^{\rm{s}}$, 
Dec.(J2000)=$+00^{\circ}04'38.2"$ with a 90\% error radius of 
1.7 arcsec \citep{Evans09b}. No source was detected by the UVOT
at the X-ray afterglow position \citep{Hoversten09}.
No prompt ground based observation was reported, probably due to 
the vicinity ($46^\circ$) to the Sun.

The data were processed with the \textsc{heasoft} 
v. 6.6.1 package and corresponding calibration files: 
standard filtering and screening criteria were applied.
BAT data analysis was based on the event data recorded from -240 s to
960 s. XRT data were acquired in  Windowed Timing (WT) mode until around 150 s; 
after that time the XRT automatically switched to the Photon Counting (PC) mode to 
follow the decaying source photon flux. Between
$\sim150$ $\sim690\,\rm{s}$ PC data were affected by pile-up: in this time 
interval an annular region of events extraction 
with the exclusion radius estimated following the
prescriptions of \cite{Moretti05} was used instead of a circular region.
The resulting $0.3-10\,\rm{keV}$ light-curve is shown in Fig. \ref{Fig:XRT}:
the chosen data binning assures a minimum signal to noise ratio (SNR) 
equal to 4; when single orbit data were not able to fulfil this requirement,
data coming from different orbits were merged to build a unique data point.

\section{Analysis and results}
\label{sec:analysis}
The BAT light curve (Fig. \ref{Fig:BAT}) shows a double-peaked
structure with $T_{90} (15-350\,\rm{keV})=24.8 \pm2.7\,\rm{s}$
\citep{Stamatikos09}. It can be fit using two \cite{Norris05} 
profiles peaking at $t_{\rm{peak,1}}=4.2\pm1.2\,\rm{s}$ and 
$t_{\rm{peak,2}}=9.3\pm1.1\,\rm{s}$;
the two structures are characterised by a $1/e$ rise and decay times
$t_{\rm{rise,1}}=2.6\pm0.5\,\rm{s}$,
$t_{\rm{decay,1}}=6.6\pm0.5\,\rm{s}$, $t_{\rm{rise,2}}=4.5\pm0.4\,\rm{s}$,
$t_{\rm{decay,2}}=8.8\pm1.3\,\rm{s}$ and 
a width $w_{1}=9.1\pm1.0\,\rm{s}$  and $w_{2}=13.3\pm1.6\,\rm{s}$. 
The amplitude is $A_1=0.46\pm0.13\,(\rm{count\,s^{-1}\,cm^{-2}})$ and 
$A_2=0.38\pm0.04\,(\rm{count\,s^{-1}\,cm^{-2}})$.
The parameters are defined following \cite{Norris05}, while their
uncertainty is computed accounting for their covariance and quoted at
68\% c.l.

The time averaged BAT spectrum can be fit by
a soft single power-law photon index $\Gamma=2.37\pm0.18$ with
a total fluence $S(15-150\,\rm{keV})=(6.2\pm1.1)\times10^{-7}$
$\rm{erg\,cm^{-2}}$ ($\chi^{2}/\rm{dof}=55.92/56$). The fluence
ratio $S(25-50\,\rm{keV})/S(50-100\,\rm{keV})=1.29\pm0.20$ (68\% c.l.) 
places GRB\,090111 at the boundary 
between X-Ray Rich (XRR) and X-Ray Flash (XRF) events according to the 
classification of \cite{Sakamoto08}.
The BAT data alone are not able to constrain the $E_{\rm{p}}$ 
parameter (peak energy of the $\nu F_{\nu}$ spectrum): however, fixing 
the low energy photon index $\alpha_{\rm{B}}$ of a Band model 
at $-1$ (typical value for both GRBs 
and XRFs, see e.g. \citealt{Sakamoto05}) we derive $E_{\rm{p}}<32\,\rm{keV}$. 
Using the $E_{\rm{p}}-\Gamma$ relation developed by \cite{Sakamoto09} 
we have $E_{\rm{p}}<27\,\rm{keV}$, in agreement with the previous result. 

The X-ray light-curve (Fig.\ref{Fig:XRT}) exhibits a steep decay which is best fit by
a simple power law with index $\alpha_{1}=5.1\pm0.2$ ($\alpha_{1}=4.6\pm0.2$)
and  $T_{0}=0$ s ($T_{0}=9.3$ s, peak time of the second prompt pulse).
This is followed by a re-brightening which dominates the light-curve between $420$ and $900\,\rm{s}$.
During this time period no detection can be reported in the 15-150 keV
energy range. After the re-brightening the light curve flattens to a simple
power law index $\alpha_{2}=0.5\pm0.2$,
while starting from $15\,\rm{ks}$ the count rate decays as 
$\alpha_{3}=1.1\pm0.3$ (Fig. \ref{Fig:XRT}). The re-brightening can be fit 
adding a \cite{Norris05} component with amplitude $A=1.53\pm0.23\,\,\rm{count\,s^{-1}}$,
start time $t_{\rm{s}}=370\,\rm{s}$ ($\chi^{2}/\rm{dof}=84.8/93$)
and rise and decay times $t_{\rm{rise}}=69.3\pm8.9\,\rm{s}$
$t_{\rm{decay}}=212.3\pm37.5\,\rm{s}$; a width $w=281.6\pm39.2\,\rm{s}$;
a peak time $t_{\rm{peak}}=472.8\pm21.0$
and an asymmetry parameter $k=0.51\pm0.04$
 of \cite{Norris05}. 
This implies a $T_{90}$ of $\sim675\,\rm{s}$.
In this time interval, the light-curve experiences a re-brightening to underlying continuum
fluence ratio $S_{\rm{reb}}/S_{\rm{cont}}\sim4.7$, while the relative 
flux variability is $\Delta F/F=14.2\pm2.1$ (where $\Delta F$ is the 
re-brightening contribution to the total flux at $t_{\rm{peak}}$ and
$F$ is underlying power-law flux at the same time). All uncertainties 
related to the light-curve fitting are quoted at 68\% c.l.

The steep decay spectrum ($77\,\rm{s}<t<150\,\rm{s}$) can
be modelled using an absorbed simple power-law with photon index
$\Gamma=5.1\pm0.4$ and neutral hydrogen column density
$N_{\rm{H},0}= (4.9\pm0.8)\times10^{21}\,\rm{cm^{-2}}$ in excess of the 
Galactic value in this direction which is $6.5\times10^{20}\,\rm{cm^{-2}}$,  
\citep{Kalberla05} ($\chi^2/\rm{dof}=68.77/49$).
While a pure black body emission model is ruled out, the addition of a 
black body component statistically improves the fit: however, the data 
are not able to simultaneously constrain the black body temperature 
and intrinsic absorption so that only rough $2\sigma$ limits can be quoted:
$0.2\,\rm{keV}<kT_{\rm{b}}<0.8\,\rm{keV}$, 
$(0.5<\rm{N_{\rm{H},0}}<5)\times 10^{22}\,\rm{cm^{-2}}$.
The X-ray data can alternatively be fit by simultaneously modelling
the Galactic and host absorption at the proper redshift. We find two
sets of allowed parameters: the first is for a close GRB with 
$\rm{N_{\rm{H},z}} = (0.63^{+0.14}_{-0.09})\times10^{22}\,\rm{cm^{-2}}$,
$z=0.5^{+0.2}_{-0.3}$ and $\Gamma=4.4\pm0.2$ 
($\chi^{2}/\rm{dof}=40.5/49$, $\rm{Pval}=80\%$). The second solution is 
for a distant and heavily absorbed GRB: 
$\rm{N_{\rm{H},z}} = (8.8^{+2.8}_{-6.1})\times10^{22}\,\rm{cm^{-2}}$,
$z=3.8^{+0.2}_{-0.3}$ and $\Gamma=4.0\pm0.2$ 
($\chi^{2}/\rm{dof}=48.6/49$, $\rm{Pval}=49\%$).
The fit is not able to constrain the redshift parameter: 
however the detection of $N_{\rm{H},z}$ in excess of the Galactic value (at $z=0$)
suggests $z<1.8$ according to the \cite{Grupe07} relation.

\begin{figure}
\centering
    \includegraphics[scale=0.525]{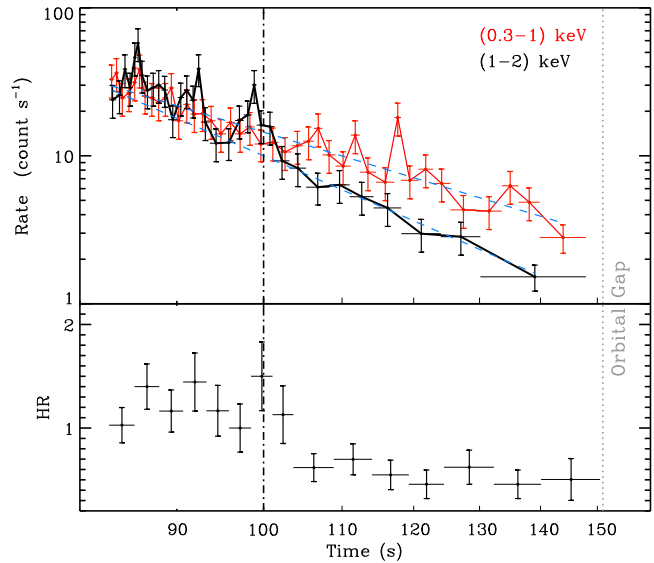}
      \caption{Upper panel: red (black) points: $0.3-1\,\rm{keV}$ 
($1-2\,\rm{keV}$) XRT light-curve rebinned at constant signal to 
noise ratio $\rm{SNR}=4$. Blue dashed lines: best fit simple power law
models. Lower panel: 
hardness ratio $\rm{HR}=(1-10)\rm{keV}/(0.3-1)\rm{keV}$ evolution with time. 
The dashed-dotted vertical line marks the beginning
of the HR decrease.}
\label{fig:hr}
\end{figure}

Spectral evolution is apparent from Fig. \ref{fig:hr},
with the $(1-2)\,\rm{keV}/(0.3-1)\,\rm{keV}$ hardness ratio starting to decrease 
$100\,\rm{s}$ after the trigger: it is interesting to note that
this corresponds to the end of a period of higher temporal variability detected in the 
$1-2\,\rm{keV}$ light-curve. This kind of variability is not seen in the
$0.3-1\,\rm{keV}$ data.
A comparison of the light-curves extracted in the two energy bands reveals
a depletion of high energy photons with time: 
while the $0.3-1\,\rm{keV}$ best fit simple power-law decay index is 
$\alpha_{1}=4.3\pm0.3$, the continuum higher energy ($1-2\,\rm{keV}$) photon flux 
decay is steeper, being modelled by $\alpha_{2}=5.6\pm0.5$.

\begin{figure}
\centering
    \includegraphics[scale=0.44]{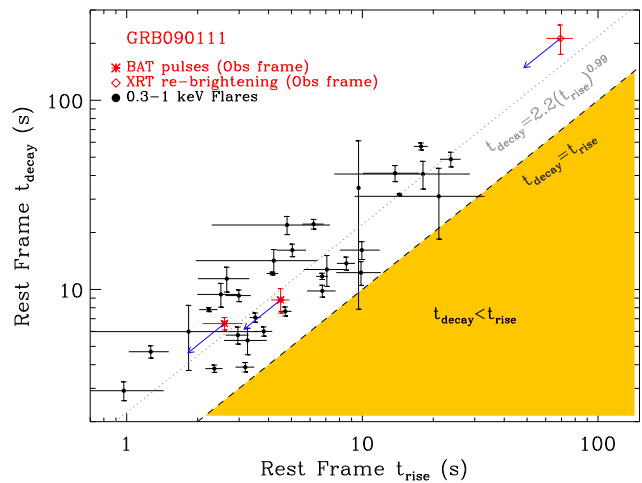}
      \caption{Decay time versus rise time for a sub-sample 
of 32 early time flares identified in the 0.3-10 keV energy range in GRBs with red-shift 
(Chincarini et al.2009 in prep.) and for GRB090111. The
blue arrows track the shift of the data when the red-shift correction is applied.  
The black dashed line
corresponds to the $t_{\rm{decay}}=t_{\rm{rise}}$ locus, while the best fit
power-law model is indicated with a grey dotted line: 
$t_{\rm{decay}}=(2.2\pm0.1)t_{\rm{rise}}^{(0.99\pm0.02)}$ ($1\sigma$ c.l.).
}
\label{Fig:trisetdecay}
\end{figure}

\begin{figure}
\centering
    \includegraphics[scale=0.43]{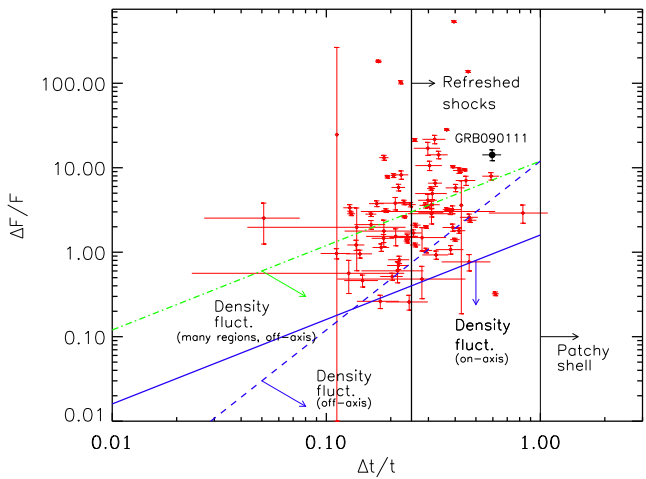}
      \caption{Relative variability flux ($\Delta F/F$) kinematically 
allowed regions as a function of relative variability time scale 
$\Delta t/t$ for a sample of 81 early ($t_{\rm{peak,obs}}<1000\,\rm{s}$) 
flares identified in 54 different GRBs (Chincarini et al. 2009 in prep.).  
The three limits shown have been computed according to Eq. 7 and A2
of Ioka et al. 2005. The position of GRB\,090111 is marked with a 
filled black dot.} 
\label{Fig:Ioka}
\end{figure}

\section{Discussion}
\label{sec:discussion}

\subsection{Unusual spectral properties}
GRB\,090111 shows a very steep ($\alpha=5.1\pm0.2$, 68\% c.l.) and soft decay 
(spectral index $\beta=4.1\pm0.4$): out of 295 GRB X-ray light-curves showing 
the canonical steep-shallow-normal decay transition analysed by \cite{Evans09a},  
only 11 (4\%) events 
are characterised by an initial power law index steeper than
the one observed in GRB\,090111. Such a high value suggests that this
is the beginning of the tail of a flare whose onset was missed by the 
XRT. The spectral analysis leads to the same conclusion:
out of 1242 time resolved XRT spectra of \emph{Swift}-GRB in the time
period April 2005 - September 2008, we found the existence of very soft absorbed simple 
power-law photon indices $\Gamma>5$ in GRB050714B, GRB050822 and GRB060512:
in each of these cases, the soft spectral emission is linked to flare activity in 
the XRT light-curve. (The three bursts also show a soft BAT prompt emission, with a time
averaged 15-150 keV photon index $\Gamma\sim2.4-2.5$). 
If this is the case, the comoving spectrum is likely to be 
a Band spectrum whose $E_{\rm{peak}}$ evolves to lower values.

Both the BAT prompt photon index steeper than 2, and the 
XRT photon index $\Gamma>4$ steeper than the typical Band low-energy photon index 
$\alpha_{\rm{B}}\sim-1$ (see e.g. \citealt{Kaneko06}, \citealt{Sakamoto05})
suggest that in both cases the observed emission is dominated by the beta portion of the
comoving Band spectrum. It is interesting to note that fixing $\alpha_B\sim-1$
in the prompt spectrum we obtain
$28\,\rm{keV}<E_{\rm{peak}}<30\,\rm{keV}$ at $3\sigma$ level for a high
energy photon index $-5<\beta_{\rm{B}}<-4$ which matches the unusual value
of the high energy photons index measured in XRT. This establishes a
spectral connection between the XRT steep decay and the prompt emission,
provided that the $E_{\rm{peak}}$ had shifted well inside the XRT energy
range by the beginning of the observation as found in other GRBs and XRFs (e.g.
GRB060614, \citealt{Mangano07a};  XRF 050416A, \citealt{Mangano07b}). At the 
same time the very soft emission observed extends
the distribution in $\beta_{\rm{B}}$ to very low values: only $\sim10\%$ of the 
spectra of 156 BATSE GRBs either have $\beta_{\rm{B}}<-4$ or do not have any high 
energy component (see e.g. \citealt{Kaneko06}).

During the steep decay spectral evolution is apparent 
(Fig. \ref{fig:hr}, lower panel). We split the steep decay phase into two time intervals, 
taking $100\,\rm{s}$ as dividing line as suggested by the hardness ratio evolution. 
A simultaneous fit of the two spectra with an absorbed cut-off power-law model (with $E_{\rm{peak}}$
as a free parameter of the fit) shows that for each  $(N_{\rm{H},z},z)$ couple
there exists a statistically acceptable solution with 
$E_{\rm{peak,1}}=1.0^{+0.2}_{-0.1}\,\rm{keV}$ and $E_{\rm{peak,2}}<0.3\,\rm{keV}$, 
where the subscripts 1 and  2
refer to the first ($t<100\,\rm{s}$) and second ($t>100\,\rm{s}$) time interval,
respectively. This suggests that the detected spectral evolution can be linked to
the evolution of the $E_{\rm{peak}}$ to lower values.
It is worth noting that the higher temporal variability 
characterising the 1-2 keV signal in the first 100 s (Fig. \ref{fig:hr}) disappears 
as the peak energy evolves outside the energy band.


\subsection{Peculiar re-brightening: a flare?}

Interpreting the X-ray re-brightening as onset of the afterglow, it is possible
to infer the initial Lorentz factor $\Gamma_{0}$ of the fireball from the 
light-curve peak time (see \citealt{Molinari07} and references therein). 
For a homogeneous surrounding medium with particle density  $n_{0}=1\,\rm{cm^{-3}}$, 
radiative efficiency $\eta=0.2$ 
we have $\Gamma_{0}\sim180 (1+z)^{3/8}(E_{\gamma}/10^{53}\rm{erg})^{1/8}$. From 
$z<1.8$ we derive an intrinsic peak energy $E_{\rm{p,i}}<84\,\rm{keV}$ and 
isotropic energy $E_{\rm{iso}}<9\times 10^{51}\,\rm{erg}$
(well within the 2$\sigma$ region of the \citealt{Amati06} relation).
This translates into a conservative upper limit $\Gamma_{0}<100$: this is lower than what is 
commonly found for normal GRBs ($\Gamma_{0}\sim500$, see e.g. \citealt{Molinari07}), 
and consistent with the less-Lorentz-boosted 
interpretation of XRRs and XRFs 
(see \citealt{Zhang07b} for a review). A similar result has been found for other XRFs:
see e.g. XRF080330 \citep{Guidorzi09}. 

In the context of off-axis emission, it is worth noting that the X-ray re-brightening 
experienced by GRB\,090111 is a sharp feature, reaching 
a flux contrast $\Delta F / F\sim 14$ during a rising time of only $\sim70\,\rm{s}$. 
\cite{Granot05} showed that both on-axis and off-axis decelerating jets 
can only produce smooth bumps in the afterglow emission.
We therefore consider this hypothesis unlikely.

A much more likely explanation is suggested by Fig. \ref{Fig:trisetdecay} where
the temporal properties of the GRB\,090111 BAT pulses and of the XRT re-brightening are 
shown to be consistent with the best fit relation found for the intrinsic 
properties of 32 0.3-10 keV early time flares (Chincarini et al. 2009, in prep.).
This fact, together with the consistency with the typical 
$t_{\rm{rise}}/t_{\rm{decay}}\sim 0.3-0.5$ \citep{Norris96} found for prompt pulses,
would suggest a common internal shock origin.

Alternatively the bump could be due to refreshed shocks 
\citep{Rees98}. Following the 
calculations of \cite{Ioka05} we plot in Fig. 
\ref{Fig:Ioka} the $\Delta F/F$ and $\Delta t/t$ values for the X-ray bump of GRB\,090111
together with the values coming from a homogeneous analysis of 82 early 
($t_{\rm{peak}}<1000\,\rm{s}$) flares identified in 54 different GRBs by Chincarini et al.
2009 in prep.: all the flares (including the GRB\,090111 bump) were fit using the 
same \cite{Norris05} profile, defining the width of each pulse as the time interval
between the $1/e$ intensity points. 
Figure \ref{Fig:Ioka} shows the kinematically allowed regions
for bumps produced by density fluctuations (\citealt{Wang00};
\citealt{Lazzati02}; \citealt{Dai02}) seen on-axis, off-axis and by many regions
according to eq. 7 and A2 in \cite{Ioka05}; bumps due to patchy shells 
(\citealt{Meszaros98}; \citealt{Kumar00b}) occupy the $\Delta t>t$ region, 
while refreshed shocks account for the $\Delta t> t/4$
area. From this figure it is apparent that the X-ray bump of GRB\,090111 lies in the refreshed 
shocks region: density fluctuations are ruled out.
    
\section{Conclusions}
\label{sec:conclusion}
GRB\,090111 shows an extra soft $\Gamma>5$ steep decay emission. This is likely due to an 
intrinsic Band spectrum whose low energy power law is missed because of the limited energy 
range of the XRT. The peak energy of the spectrum evolves through the XRT band producing a 
softening trend testified by the different light-curve decay behaviours in different energy 
bands. It's interesting to note that the period of higher temporal variability in the
1-2 keV light-curve ends when the $E_{\rm{peak}}$ shifts outside the energy band.
The steep decay is followed by an X-ray re-brightening whose peculiar temporal properties 
made it worth a detailed study. While the temporal properties of the re-brightening are
consistent with an internal origin, with $\Delta t/t \sim 0.6$ and $\Delta F/ F\sim14$ the bump
lies in the refreshed shocks region of Fig. \ref{Fig:Ioka}. Density fluctuations are
ruled out. 
Finally, with a fluence ratio $S(25-50\,\rm{keV})/S(50-100\,\rm{keV})=1.29\pm0.20$ (68\% c.l.)   
we propose this event to be classified as XRR\,090111.

\section*{Acknowledgements}
We thank the referee for constructive criticism.
This work is supported by ASI grant SWIFT I/011/07/0, by  the 
Ministry of University and Research of Italy (PRIN MIUR 2007TNYZXL), by
MAE and by the University of Milano Bicocca (Italy).

\newpage

\bsp

\label{lastpage}

\end{document}